%% file: 0-main.tex
\documentclass[sigconf,screen,nonacm,review=false]{acmart}
\usepackage{booktabs} 
\usepackage{multirow}

\usepackage{color}

\usepackage{subfigure} 
\usepackage{array}
\usepackage{breqn}
\usepackage{{inputenc}}
\usepackage{makecell}
\usepackage{balance}
\usepackage{multirow,tabularx}
\usepackage{longtable}
\usepackage{booktabs,longtable}
\usepackage{hhline}
\usepackage[flushleft]{threeparttable}
\usepackage[ruled, lined, linesnumbered, commentsnumbered, longend]{algorithm2e}
\usepackage{epsfig}
\usepackage{graphicx}
\usepackage{epstopdf}
\usepackage{thmtools}
\usepackage{enumitem}
\usepackage{verbatim}
\usepackage{amsfonts}
\usepackage{hyperref}
\usepackage{arydshln}
\usepackage{caption}
\usepackage{subcaption}
\usepackage{svg}
\usepackage{amsmath}

\usepackage{tcolorbox}
\usepackage{xcolor}
\usepackage{svg}
\usepackage{bm}
\usepackage{afterpage}
\usepackage{mdframed}
\tcbuselibrary{skins, breakable}

\definecolor{background_b}{HTML}{DFEEEA}
\definecolor{frame_b}{HTML}{80A6E2}
\definecolor{myblue}{HTML}{CF553D}
\definecolor{myblue2}{HTML}{4472C4}

\newsavebox{\InterviewCase}

\newcommand{\ie}{\emph{i.e., }}
\newcommand{\eg}{\emph{e.g., }}

\newcommand{\etc}{\emph{etc.}}
\newcommand{\wrt}{\emph{w.r.t. }}

\newcommand{\aka}{\emph{aka. }}

\newcommand{\seqemb}[1]{\texttt{{\color{myblue}{#1}}}}

\newcommand{\be}{\mathbf{e}}
\newcommand{\bs}{\mathbf{s}}
\newcommand{\bh}{\mathbf{h}}
\newcommand{\bE}{\mathbf{E}}
\newcommand{\bX}{\mathbf{X}}

\AtBeginDocument{%
  \providecommand\BibTeX{{%
    \normalfont B\kern-0.5em{\scshape i\kern-0.25em b}\kern-0.8em\TeX}}}

\setcopyright{acmcopyright}
\copyrightyear{2018}
\acmYear{2018}
\acmDOI{XXXXXXX.XXXXXXX}

\acmConference[Conference acronym 'XX]{Make sure to enter the correct
  conference title from your rights confirmation emai}{June 03--05,
  2018}{Woodstock, NY}
\acmPrice{15.00}
\acmISBN{978-1-4503-XXXX-X/18/06}

\begin{document}

\title{Large Language Model Can Interpret Latent Space \\ of Sequential Recommender}



\author{Zhengyi Yang}
\affiliation{%
  \institution{University of Science and Technology of China}
  \country{Hefei, China}}
\email{yangzhy@mail.ustc.edu.cn}

\author{Jiancan Wu}
\affiliation{%
  \institution{University of Science and Technology of China}
  \country{Hefei, China}}
\email{wujcan@gmail.com}

\author{Yanchen Luo}
\affiliation{%
  \institution{University of Science and Technology of China}
  \country{Hefei, China}}
\email{luoyanchen@mail.ustc.edu.cn}

\author{Jizhi Zhang}
\affiliation{%
  \institution{University of Science and Technology of China}
  \country{Hefei, China}}
\email{cdzhangjizhi@mail.ustc.edu.cn}

\author{Yancheng Yuan}
\affiliation{%
  \institution{The Hong Kong Polytechnic University}
  \country{Hong Kong, China}}
\email{yancheng.yuan@polyu.edu.hk}

\author{An Zhang}
\affiliation{%
  \institution{National University of Singapore}
  \country{Singapore, Singapore}}
\email{anzhang@u.nus.edu}

\author{Xiang Wang}
\affiliation{%
  \institution{University of Science and Technology of China}
  \country{Hefei, China}}
\email{xiangwang1223@gmail.com}

\author{Xiangnan He}
\affiliation{%
  \institution{University of Science and Technology of China}
  \country{Hefei, China}}
\email{xiangnanhe@gmail.com}

\renewcommand{\shortauthors}{Zhengyi Yang, et al.}

\input{chapters/0_abs}
\ccsdesc[500]{Information systems~Recommender systems}
\ccsdesc[500]{Information systems~Retrieval models and ranking}
\keywords{Sequential Recommendation, Large Language Models}



\maketitle

\input{chapters/1_intro-new}

\input{chapters/2_related}

\input{chapters/3_method}

\input{chapters/4_experiment}

\input{chapters/5_conclusion}

\bibliographystyle{ACM-Reference-Format}
\bibliography{9_reference}

\end{document}

%% file: chapters/0_abs.tex
\begin{abstract}
    Sequential recommendation is to predict the next item of interest for a user, based on her/his interaction history with previous items. In conventional sequential recommenders, a common approach is to model item sequences using discrete IDs, learning representations that encode sequential behaviors and reflect user preferences.
    Inspired by recent success in empowering large language models (LLMs) to understand and reason over diverse modality data (\eg image, audio, 3D points), a compelling research question arises: ``Can LLMs understand and work with hidden representations from ID-based sequential recommenders?''.
    To answer this, we propose a simple framework, RecInterpreter, which examines the capacity of open-source LLMs to decipher the representation space of sequential recommenders.
    Specifically, with the multimodal pairs (\ie representations of interaction sequence and text narrations), RecInterpreter first uses a lightweight adapter to map the representations into the token embedding space of the LLM.
    Subsequently, it constructs a sequence-recovery prompt that encourages the LLM to generate textual descriptions for items within the interaction sequence.
    Taking a step further, we propose a sequence-residual prompt instead, which guides the LLM in identifying the residual item by contrasting the representations before and after integrating this residual into the existing sequence.
    Empirical results showcase that our RecInterpreter enhances the exemplar LLM, LLaMA, to understand hidden representations from ID-based sequential recommenders, especially when guided by our sequence-residual prompts.
    Furthermore, RecInterpreter enables LLaMA to instantiate the oracle items generated by generative recommenders like DreamRec, concreting the item a user would ideally like to interact with next.
    Codes are available at \url{https://github.com/YangZhengyi98/RecInterpreter}.

\end{abstract}

%% file: chapters/1_intro-new.tex
\section{Introduction}


\begin{figure*}
    \centering
    \subfigure[Sequential Recommendation (SeqRec)]{
        \includegraphics[height=6cm, trim={1.4cm 0 1.4cm 0}, clip]{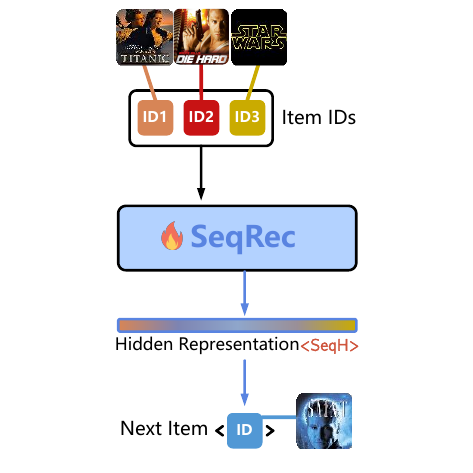}
        \label{fig:intro-1}
    }
    \subfigure[LLM4Rec]{
        \includegraphics[height=6cm, trim={1cm 0 1cm 0}, clip]{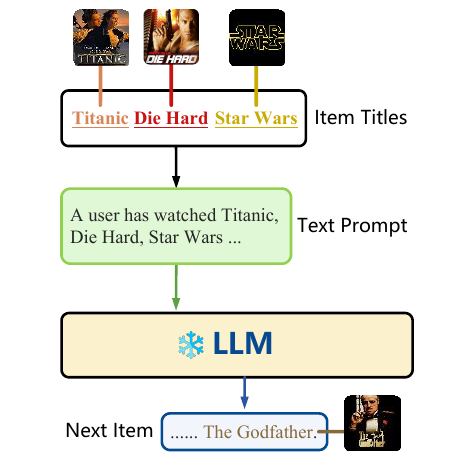}
        \label{fig:intro-2}
    }
    \subfigure[Rec Interpreter]{
        \includegraphics[height=6cm]{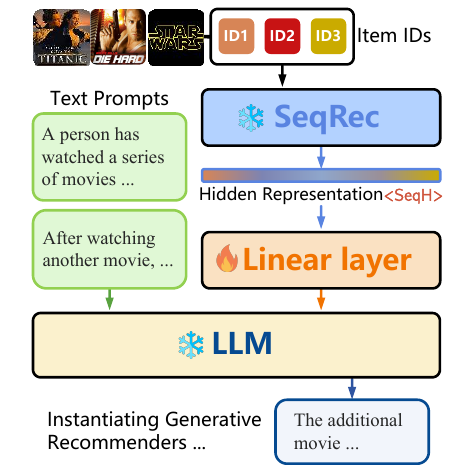}
        \label{fig:intro-3}
    }
    \vspace{-0.3cm}
    \caption{Comparision of conventional sequential recommendation, LLM4Rec and RecInterpreter. Flame \includegraphics[width=0.015\textwidth]{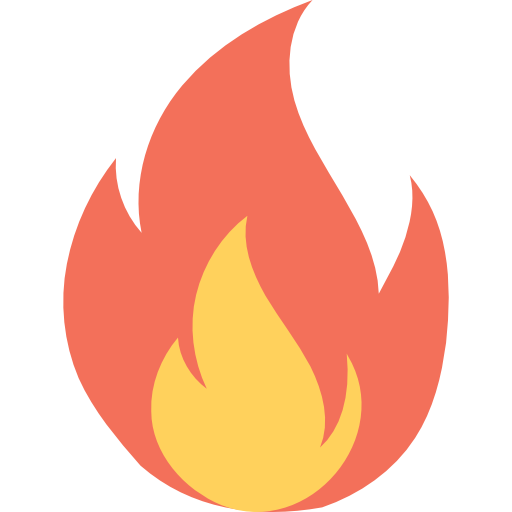} denotes tunable modules, while snowflake \includegraphics[width=0.015\textwidth]{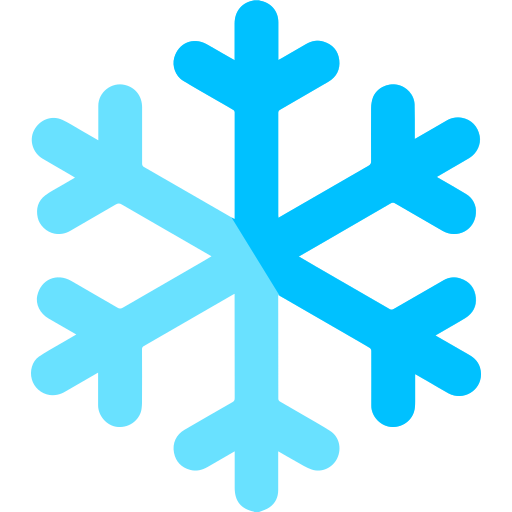} indicates frozen modules.}
    \vspace{-0.3cm}
    \label{fig:intro}
\end{figure*}

Sequential recommendation --- predicting the next item of interest based on a sequence of items that a user interacted with before --- has been a fundamental task in both academia and industry \cite{GRU4Rec,SASRec,Youtube}. 
Scrutinizing leading sequential recommenders \cite{GRU4Rec,Caser,SASRec,DROS,Google}, we can summarize a typical pipeline: 1) assign discrete IDs to items and initialize learnable vectors (\aka item embeddings) to represent different items, and 2) learn the hidden representation based on each sequence of item embeddings, as Figure \ref{fig:intro-1} shows.
Such representations, derived from the ID-modeling paradigm, are able to encode sequential patterns of user behaviors, having greatly facilitated the next-item recommendation.

With the meteoric rise of Large Language Models (LLMs) (\eg GPT4 \cite{GPT4}, LLaMA \cite{llama}), aligning diverse modalities --- such as images, audios, and 3D points --- with text can empower LLMs to understand and reason about other modalities \cite{flamingo, minigpt, audio-llm, video-llm, palm-e, 3d-llm, point-llm}.
Central to such an alignment is transforming the hidden representations from the modality-specific encoders (\eg images encoded by ViT \cite{vit} or Stable Diffusion \cite{stable-diffusion}, audios encoded by HiFiGAN \cite{hifigan}) into the text token embeddings of an LLM \cite{minigpt, audio-llm}. 
This allows for the LLM to reason over the input modality and generate the textual responses correspondingly.
Although multi-modal comprehension is becoming a focal point of LLMs, the capability to interpret hidden representations from sequential recommenders remains mostly unexplored.
This is largely due to the current LLMs-for-Recommendation (LLM4Rec) studies \cite{P5, rec_survey_1, rec_survey_2, rec_survey_3, bigrec, ai2023information}, which primarily focus on recasting the user-item interactions as text prompts and feeding them into LLMs for recommendation or reranking, as Figure \ref{fig:intro-2} illustrates.
However, this paradigm shields LLMs from accessing or deciphering the hidden representations from recommender models.

Naturally, an compelling research question arises: ``\textbf{Can LLMs Understand Representations from Recommenders?}''
To answer this, we propose a simple framework, \textbf{RecInterpreter}, which examines the capacity of open-source LLMs to decipher the representation space of sequential recommenders.
Here we select LLaMA-7B \cite{llama} as a prime example of open-source LLMs, which offer access to its hidden states and support backpropagation.
In terms of sequential recommenders, we harness the representative models trained solely on item ID sequences, including GRU4Rec \cite{GRU4Rec}, Caser \cite{Caser}, SASRec \cite{SASRec}, and DreamRec \cite{dreamrec}.
Having the LLM and recommender frozen, one straightforward solution to bridge their gap is the alignment training \cite{flamingo,minigpt,anymal} with the paired multimodal data (\ie representations of item ID sequences and text narrations).
Following the leading alignment strategies \cite{minigpt, palm-e, anymal}, RecInterpreter has two key components: 1) train a lightweight adapter to map the recommendation representations into the token embedding space of the LLM, and 2) inject these recommendation-specific tokens into a text prompt to ask the LLM for a textual elucidation.
Next, we will elaborate on these components.

Specifically, as a bridge, the adapter fuses the spaces of the LLM and recommender into a joint token embedding space, wherein tokens represent both text and user behavior.
Moreover, we simply set it as a single linear projection layer to train, where the model parameters of the recommenders and LLaMA are frozen.
This lightweight design not only reaches convergence faster than training from scratch, but also inherits the reasoning capabilities of the LLM.

Having the recommendation-specific tokens, we first propose the sequence-recovery prompt, which tries to recover the whole item sequence. 
Here is an example of the prompt in the movie recommendation scenario:
\begin{tcolorbox}[boxrule=0.5pt, title={Sequence-Recovery Prompt Example}]
``A user has watched a series of movies, which can be represented as \seqemb{<SeqH>}. What movies has the user watched?''
\end{tcolorbox}
\noindent where $\seqemb{<SeqH>}$ is the hidden representation of watching history encoded by a sequential recommender (\eg the hidden representation in Figure \ref{fig:intro}).  
While we empirically show that LLaMA could understand some interactions from the hidden representation, it is hard to recover all the interactions, since the hidden representation is highly compressed.
To this end, we carefully craft a sequence-residual prompt tailored for sequential recommenders. This prompt is designed to guide LLaMA in identifying the residual item by comparing the representations before and after the sequence incorporates said residual.
\begin{tcolorbox}[boxrule=0.5pt, title={Sequence-Residual Prompt Example}]
``A user has watched a series of movies, which can be represented as \seqemb{<SeqH1>}. After watching another movie, the watching history can be represented as \seqemb{<SeqH2>}. What is the additional movie the user watched?''
\end{tcolorbox}
\noindent where \seqemb{<SeqH1>} and \seqemb{<SeqH2>} are the hidden representations before and after the sequence integrates with the residual.

Surprisingly, our empirical evaluations show that LLaMA exhibits a significant aptitude for deciphering the representations from sequential recommenders, especially following our instructions.
When presented with an item ID sequence and another extended by a target item ID, LLaMa can clearly tell their representation difference and yield the textual description of the target item.
Consequently, we may safely reach the conclusion that LLMs could be inspired to understand the representation space of sequential recommenders, which inherently encapsulate rich patterns of user behaviors. 
Moreover, since the linear projection is the only tunable component, it is affordable for online service providers to interpret their own recommenders with LLMs, which is flexible for them to investigate further how to utilize LLMs in their platforms.

Furthermore, another interesting research question emerges: ``\textbf{Can LLMs Instantiate the Generated Items from Generative Recommenders?}''.
Our RecInterpreter presents a straightforward solution to decode the generated results of generative recommender systems.
Take DreamRec \cite{dreamrec} --- one of the latest in the generative recommendation --- as an example. Here, the generated oracle item is encoded as vector representation, which lacks explicit interpretation.
Using our RecInterpreter framework that can differentiate between representations before and after a user's engagement with a new item, we could append the generated oracle item at the end of the interaction sequence, and let LLaMA interpret the oracle item with a text description.
Experiments demonstrate that our RecInterpreter framework can decode reasonable oracle items not limiting in the candidate set, which completes the full promise of DreamRec as a generative recommender.

\input{chapters/99_framework-sequence-recovery}

%% file: chapters/99_framework-sequence-recovery.tex
\begin{figure*}[t]
	\centering
	\small	
	\includegraphics[width=0.98\textwidth]{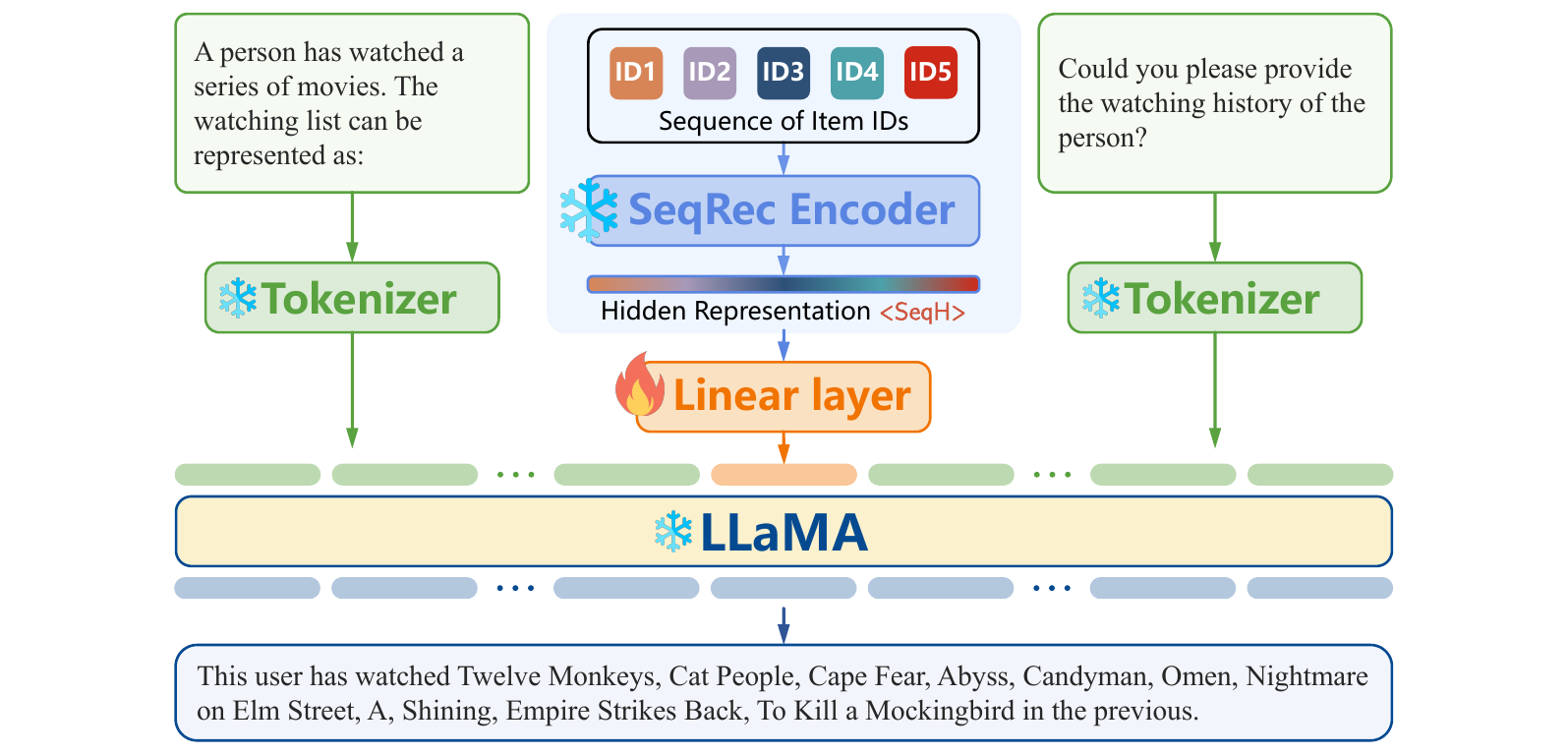}
        \vspace{-0.3cm}
	\caption{Illustration of the sequence recovery framework. We provide the task-specific textual prompts and the hidden representation of the interaction sequence projected by a linear layer, targeting at inspiring LLaMA to recover the interactions with a textual response.  Flame \includegraphics[width=0.015\textwidth]{figures/Assets/flame.png} denotes tunable modules, while snowflake \includegraphics[width=0.015\textwidth]{figures/Assets/snowflake.png} indicates frozen modules.}
	\label{fig:method_seqnence-recovery}
	\vspace{-0.3cm}
\end{figure*}

%% file: chapters/2_related.tex
\section{Related Work}

This section reviews the work on multimodal language models, and then discusses the work on sequential recommendation, especially the integration of LLMs.

\subsection{Multimodal Language Models}
Recent advances in LLMs have demonstrated remarkable few/zero-shot reasoning capabilities in Neural Language Processing tasks \cite{GPT4, InstructionTuning, zhao2023survey}.
Meanwhile, different modalities (including vision, video, audio and \etc), have been evolving their models rapidly to better accommodate different tasks \cite{vit, stable-diffusion, hifigan}.
More recently, researchers find that models of different modalities can be unified with  LLMs by making the hidden representations perceivable for LLMs, leading to a promising direction, multimodal language models \cite{flamingo, minigpt, audio-llm, video-llm, nextgpt}.

Along this research line, the pioneer work, Flamingao \cite{flamingo} demonstrates that the vision encoder NFNet \cite{nfnet} could be understood by LLMs through inserting tunable gated cross-attention dense blocks among the layers of LLMs, which has been proven effective in GPT-4Vision \cite{gpt4vision}.
MiniGPT4 \cite{minigpt} further shows that a single linear layer is enough to make LLaMA \cite{llama} to interpret hidden representations encoded by ViT \cite{vit}.
Similarly, TANGO \cite{audio-llm} suggests that the audio backbone model HiFiGAN \cite{hifigan} can be unified by LLMs.
Besides, Video-ChatGPT \cite{video-llm} and VideoChat \cite{videochat} imply that pre-trained video encoders are also perceivable for LLMs.

\subsection{Sequential Recommendation}
Sequential recommendation aims at inferring users' preferences based on their interaction sequences.
Previous work has explored encoding the interaction sequences with different model architectures, such as Recurrent Neural Network (RNN) \cite{GRU4Rec}, Convolutional Neural Network (CNN) \cite{Caser},
and Transformer encoder \cite{SASRec}. 
Moreover, recent work also designs auxiliary learning tasks like causal inference \cite{PDA, ASR}, data augmentation \cite{CL4SRec}, and robust learning \cite{DROS}.

Advances in LLMs have drawn increasing research attention to leveraging LLMs for sequential recommendation.
{Some research directly employs In-Context Learning to assess the recommendation performance of LLM or enhance traditional recommendations, thereby measuring the recommendation capabilities of LLMs~\cite{ICL_1, fairrec, ICL_6, Liu2023IsCA}.
Furthermore, some recent studies \cite{tallrec,bigrec} are concerned that LLMs lack recommendation-specific knowledge in the pre-training phase, thus proposing to leverage LLMs to implement specific tuning techniques to enhance their recommendation performance~\cite{tuning_1, tuning_3, tuning_5, bigrec, tallrec}.
}
However, these approaches mostly shield LLMs from accessing or deciphering the hidden representations from frozen recommender models.
Therefore, inspiring LLMs to interpret hidden representations from sequential recommenders remains largely unexplored.

%% file: chapters/3_method.tex
\section{Inspire LLMs to understand sequential recommenders}

In this section, we outline our RecInterpreter framework to harness the capabilities of LLMs for comprehending traditional sequential recommenders.
We first introduce a sequence-recovery task, aiming to empower LLaMA to reconstruct the items in an interaction sequence based solely on its hidden representation.
Taking a step further, we propose a novel sequence-residual task, which guides LLaMA to pinpoint the residual item by contrasting the hidden representations before and after integrating this residual into the existing sequence.
Finally, we highlight how RecInterpreter can explicitly decode the embeddings of an unseen oracle item, when employed within generative recommender settings.

\subsection{Sequence-Recovery Prompting}
To validate LLaMA's capability in understanding sequential recommenders, we draw inspiration from prior multi-modal alignment studies \cite{flamingo, minigpt, video-chatgpt, palm-e} and propose a sequence recovery task. 
This objective is to encourage LLaMA to reconstruct the items in an interaction sequence in text, based solely on its hidden representation, as depicted in Figure  \ref{fig:method_seqnence-recovery}.
Next, we will elaborate on our steps.

\vspace{5pt}\noindent\textbf{Sequence Encoding via Sequential Recommenders.}
For an interaction sequence $\bs = [s_1, s_2, \ldots, s_m]$ that involves the consumed items, we employ a well-trained sequential recommender, such as SASRec and DreamRec, to derive the hidden representation of the sequence.
Formally, this sequence encoding process can be formulated as follows:
The interaction sequence $\bs$ is first vectorized as $\bE_{\bs} =[\be_{s_1}, \be_{s_2}, \ldots, \be_{s_m}]$ by the sequential recommender, and we can acquire the hidden representation through:
\begin{equation}
    \bh_{\bs} = \textbf{Seq-Enc}(\bE_{\bs}),
\end{equation}
where $\textbf{Seq-Enc}(\cdot)$ is the sequence encoder of conventional sequential recommender, and $\bh_\bs \in \mathbb{R}^d$ is the $d$-dimentional representation of sequence $s$ (\eg $d$ is set as 64 or 256 in SASRec).

\vspace{5pt}\noindent\textbf{Representation Adaptaion vis Lightweight Adapter.}
We train a lightweight adapter to project the hidden representation $\bh_\bs$ into the text token embedding space of LLaMA.
Here we implement the adapter as a linear projection layer, whose design ensures that the input dimension aligns with $d$, while the output dimension matches LLaMA's token embedding size (\ie 4096).
Thus the hidden representation is transformed as:
\begin{equation}
    \widetilde{\bh}_\bs = \textbf{Linear-Proj}_\theta(\bh_\bs).
\end{equation}
In this way, the adapter serves as a bridge, integrating the spaces of LLaAM and the recommender system.
This leads to a unified token embedding space, where tokens can signify either textual content or user interactions.
The deeper exploration of such adapters, such as Q-former \cite{blip-2}, is an avenue we plan to explore in future work.

\vspace{5pt}\noindent\textbf{Prompt Design for the Adapter Training.}
Here we design the sequence-recovery prompts, which are composed of text tokens interleaved with the projected sequence representation $\widetilde{\bh}_\bs$.
Here is an example of the sequence-recovery prompt in the movie recommendation scenario:
\begin{tcolorbox}[boxrule=0.5pt, left=0pt, right=0pt, top=2.5pt, bottom=2.5pt, title={Sequence-Recovery Prompt}]
    \centering
    \begin{tabular}{cl}
         \makecell[c]{Input \\ Prompt}& \makecell*[{{p{5.3cm}}}]{{A person has watched a series of movies. 
         The watching list can be represented as:} \seqemb{<SeqH>}.
         {Describe this watching history of the person in detail.}}\\
         \specialrule{0em}{2.5pt}{2.5pt}
         \hline
         \specialrule{0em}{2.5pt}{2.5pt}
         \makecell[c]{Target \\ Response}& \makecell*[{{p{5.3cm}}}]{{This user has watched \textcolor{myblue}{Twelve Monkeys, Cat People, Cape Fear, Abyss, Candyman, Omen, Nightmare on Elm Street, Shining, Empire Strikes Back, To Kill a Mockingbird} in the previous.}}
    \end{tabular}
\end{tcolorbox}
\noindent where \seqemb{<SeqH>} is the projected hidden representation (\ie $\widetilde{\bh}_\bs$).

It is worth noting that the prompt involves two key components: 1) the input prompt, which contains the projected hidden representation of the sequence $\widetilde{\bh}_\bs$; and 2) the target response, which offers the detailed textual narration of $\widetilde{\bh}_\bs$.
Within the autoregressive framework of LLaMA, we calculate the training objective by regressing the target prompt $\bX_{Target}$ based on the condition of the input prompt $\bX_{Input}$ \cite{llama}:
\begin{equation}
    p(\bX_{Target} | \bX_{Input}) = \prod_{i=1}^N p(\bX_{Target}^{i} | \bX_{Input}, \bX_{Target}^{[1:i-1]}),
\end{equation}
where $N$ is the number of tokens in the target prompt.

During the training phase, we provide both the input prompt and the target response with the objective of learning to generate descriptions for the projected sequence embedding $\widetilde{\bh}_\bs$.
During the inference phase, we provide only the input prompt containing the projected sequence embedding $\widetilde{\bh}_\bs$, and acquire the output text as the understanding of $\widetilde{\bh}_\bs$.

\input{chapters/99_framework-sequence-residual}

\subsection{Sequence-Residual Prompting} \label{subsec: method2}

It is challenging for LLaMA to understand all items from a simple hidden representation of the interaction sequence, since the datasets of recommendation are usually very sparse.
Although we empirically show that LLaMA can understand the interactions to a large extent under the sequence-recovery framework, we would also like to refine the framework to encourage LLaMA to understand sequential recommender more delicately. 
Drawing inspiration from Flamingo \cite{flamingo}, which suggests that LLMs could better process images if the hidden representations of two similar images are provided at the same time with their differences, we propose to inspire LLaMA to understand sequential recommenders by identifying the residual item based on hidden representations before and after a sequence integrates the residual, as illustrated in Figure \ref{fig:method_recoverOne}.
Then we elaborate on the sequence-residual prompting step by step:

\vspace{5pt}\noindent\textbf{Sequence Encoding via Sequential recommenders.}
Given an interaction sequence $\bs = [s_1, s_2, \ldots, s_m]$, we could design a circumstance, that a user has interacted with $[s_1, s_2, \ldots, s_{m-1}]$ and then interacts with a residual item $s_m$. 
The sequential recommender could encode $\bs^1 = [s_1, s_2, \ldots, s_{m-1}]$ and $\bs^2 = [s_1, s_2, \ldots, s_{m}]$ as ${\bh_s^1}$ and ${\bh_s^2}$ respectively:
\begin{equation}
    \bh_{\bs^1} = \textbf{Seq-Enc}(\bE^1) \quad \textbf{and} \quad \bh_{\bs^2} = \textbf{Seq-Enc}(\bE^2),
\end{equation}
where $\bE^1$ and $\bE^2$ are the vactorized sequence of $\bs^1$ and $\bs^2$.

\vspace{5pt}\noindent\textbf{Representation Adaptaion vis Lightweight Adapter.}
We also employ a linear projection layer as the lightweight adapter, which could project ${\bh_{\bs^1}}$ and ${\bh_{\bs^2}}$ to be $\widetilde{\bh}_{\bs^1}$ and $\widetilde{\bh}_{\bs^2}$:
\begin{equation}
    \widetilde{\bh}_{\bs^1} = \textbf{Linear-Proj}_\theta(\bh_{\bs^1}) \quad \textbf{and} \quad \widetilde{\bh}_{\bs^2} = \textbf{Linear-Proj}_\theta(\bh_{\bs^2}),
\end{equation}
where the parameters of linear layer are shared by $\bh_{\bs^1}$ and $\bh_{\bs^2}$.

\vspace{5pt}\noindent\textbf{Prompt Design for the Adapter Training.}
Here we design more delicate sequence-residual prompts, which inspire LLaMA to identify the residual item $s_m$ by comparing $\widetilde{\bh}_{\bs^1}$ and $\widetilde{\bh}_{\bs^2}$.
Here is an example of the sequence-residual prompt in the movie recommendation scenario:
\begin{tcolorbox}[boxrule=0.5pt, left=0pt, right=0pt, top=2.5pt, bottom=2.5pt, title={Sequence-Residual Prompt}]
    \centering
    \begin{tabular}{cl}
         \makecell[c]{Input \\ Prompt}& \makecell*[{{p{5.3cm}}}]{{A person has watched a series of movies. 
         The watching list can be represented as List1:} \seqemb{<SeqH1>}.
         {After watching another movie, the watching list can further be represented as List2:} \seqemb{<SeqH2>}. {What is the movie in List2 but not in List1?}}\\ 
         \specialrule{0em}{2.5pt}{2.5pt}
         \hline
         \specialrule{0em}{2.5pt}{2.5pt}
         \makecell[c]{Target \\ Response}& \makecell*[{{p{5.3cm}}}]{{This user watched movie \textcolor{myblue}{Twelve Monkeys} in List2 but not in List1.}} \\ 
    \end{tabular}
\end{tcolorbox}
\noindent where $\seqemb{<SeqH1>}$ and $\seqemb{<SeqH2>}$ would be replaced with $\widetilde{\bh}_{\bs^1}$ and $\widetilde{\bh}_{\bs^2}$, and \textcolor{myblue}{Twelve Monkeys} is the residual item in the example.

Similar to the sequence-recovery prompting, we provide both the input prompt and target response during the training phase, and only the input prompt during the inference phase.

\subsection{Instantiate Oracle Items}
Having shown that LLaMA can be inspired to identify the residual item by comparing two hidden representations of designed sequences, we then elaborate on how to benefit generative recommender with this sequence-residual prompting.

\vspace{5pt}\noindent\textbf{Brief on DreamRec.}
In DreamRec \cite{dreamrec}, one of the latest in generative recommendation, an oracle item could be generated through the guided diffusion process. 
However, the oracle item is represented as a hidden vector without explicit interpretation, thus the completion of the recommendation task is compromised by finding the nearest items of the oracle item in the candidate set, which fails to achieve the full promise of DreamRec to generative items beyond the candidates \cite{dreamrec}.
Drawing inspiration from the proposed sequence-residual prompting, we could let LLaMA provide the description of the oracle item generated by DreamRec, thus directly acquiring the recommendation results.

\vspace{5pt}\noindent\textbf{Construct Sequence-Residual Task with Oracle Item.}
Let $\bs = [s_1, s_2, \ldots, s_n]$ be a historical interaction sequence of a user, and DreamRec can generate the vector representation of the corresponding oracle item as $\be_{s^*}$.
As described in Section \ref{subsec: method2}, the well-trained sequence-residual framework could identify the residual item between two hidden representations before and after the sequence interacts with a new item.
Therefore, we assume that the user would interact with the oracle item and construct the vectorized sequence $\bE^* = [\be_{s_1}, \be_{s_2}, \ldots, \be_{s_n}, \be_{s^*}]$.
Applying the sequential encoder of DreamRec and the linear projection adapter, we have:
\begin{equation}
    \bh_\bs = \textbf{Seq-Enc}(\bE) \quad \text{and} \quad \bh_{\bs^*} = \textbf{Seq-Enc}(\bE^*),
\end{equation}
and:
\begin{equation}
    \widetilde{\bh}_{\bs} = \textbf{Linear-Proj}_\theta(\bh_\bs) \quad \text{and} \quad \widetilde{\bh}_{\bs^*} = \textbf{Linear-Proj}_\theta(\bh_{\bs^*}).
\end{equation}
To this end, LLaMA could identify the oracle item $\be_{s^*}$ with textual descriptions by comparing $\widetilde{\bh}_{\bs}$ and $\widetilde{\bh}_{\bs^*}$ with the sequence-residual prompting framework.

\vspace{5pt}\noindent\textbf{Training and Inference.}
The training phase remains the same as the sequence-residual prompting framework, \ie we utilize the sequences in the dataset to construct the contrastive hidden representations pairs for training.
During the inference phase, we would feed $\widetilde{\bh}_\bs$ and $\widetilde{\bh}_{\bs^*}$ into LLaMA, and then LLaMA could respond with a textual description about the oracle item.
Therefore, we can complete the explicit decoding of the generated oracle items, which has not been achieved by DreamRec.

%% file: chapters/99_framework-sequence-residual.tex
\begin{figure*}[t]
	\centering
	\small	
	\includegraphics[width=0.98\textwidth]{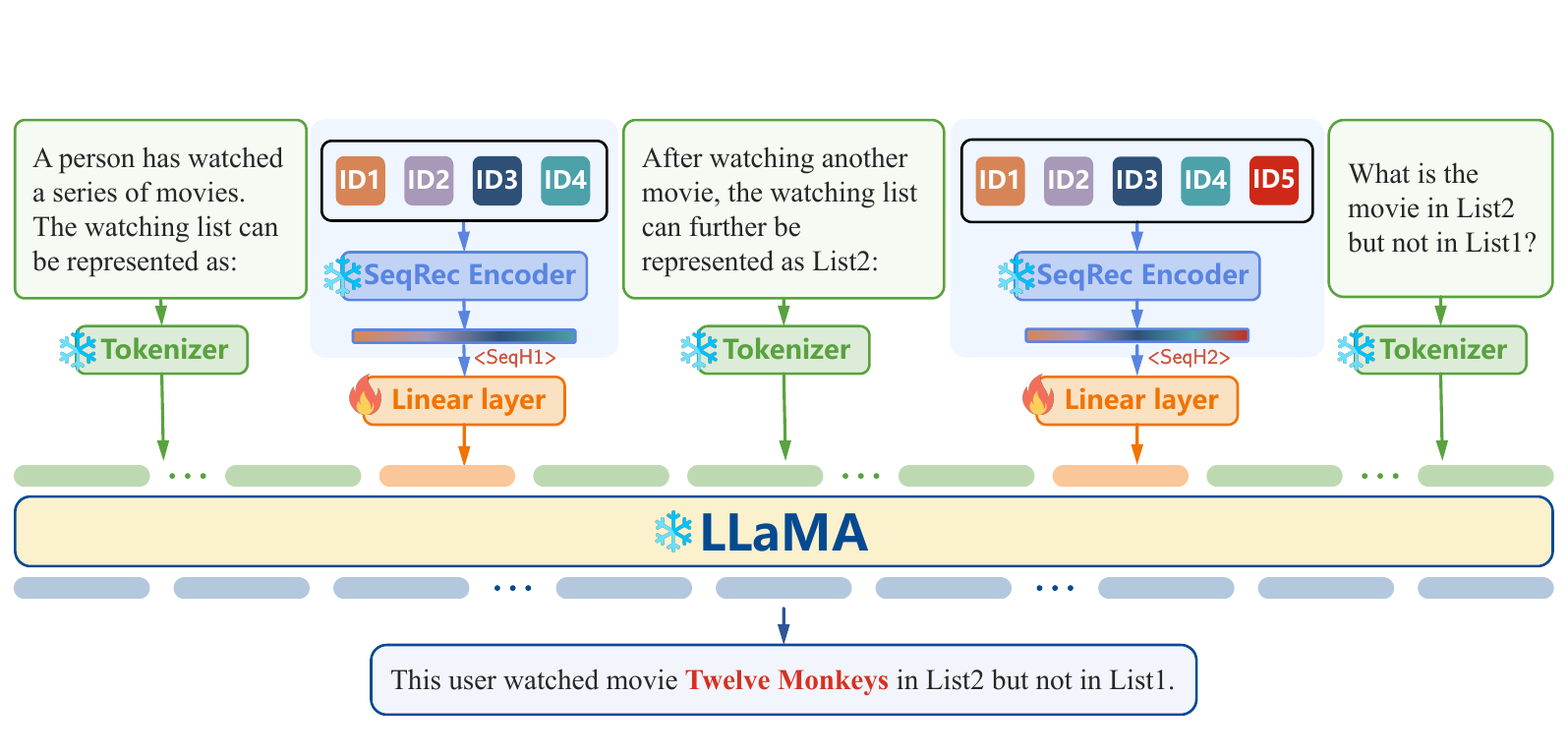}
        \vspace{-0.3cm}
	\caption{Illustration of the sequence residual framework. We provide the task-specific textual prompts and the hidden representations before and after the sequence incorporates a residual item. The two hidden representations are projected by a shared linear Layer. Flame \includegraphics[width=0.015\textwidth]{figures/Assets/flame.png} denotes tunable modules, while snowflake \includegraphics[width=0.015\textwidth]{figures/Assets/snowflake.png} indicates frozen modules.}
	\label{fig:method_recoverOne}
	\vspace{-0.3cm}
\end{figure*}

%% file: chapters/4_experiment.tex
\section{Experiment}
In this section, we conduct experiments to demonstrate our approach to inspire LLMs to understand sequential recommenders through interpreting the hidden representations.
Then we show how to facilitate generative recommendation by providing a textual description of generated items.

\begin{figure*}[t]
	\centering
	\small	
	\includegraphics[width=0.98\textwidth]{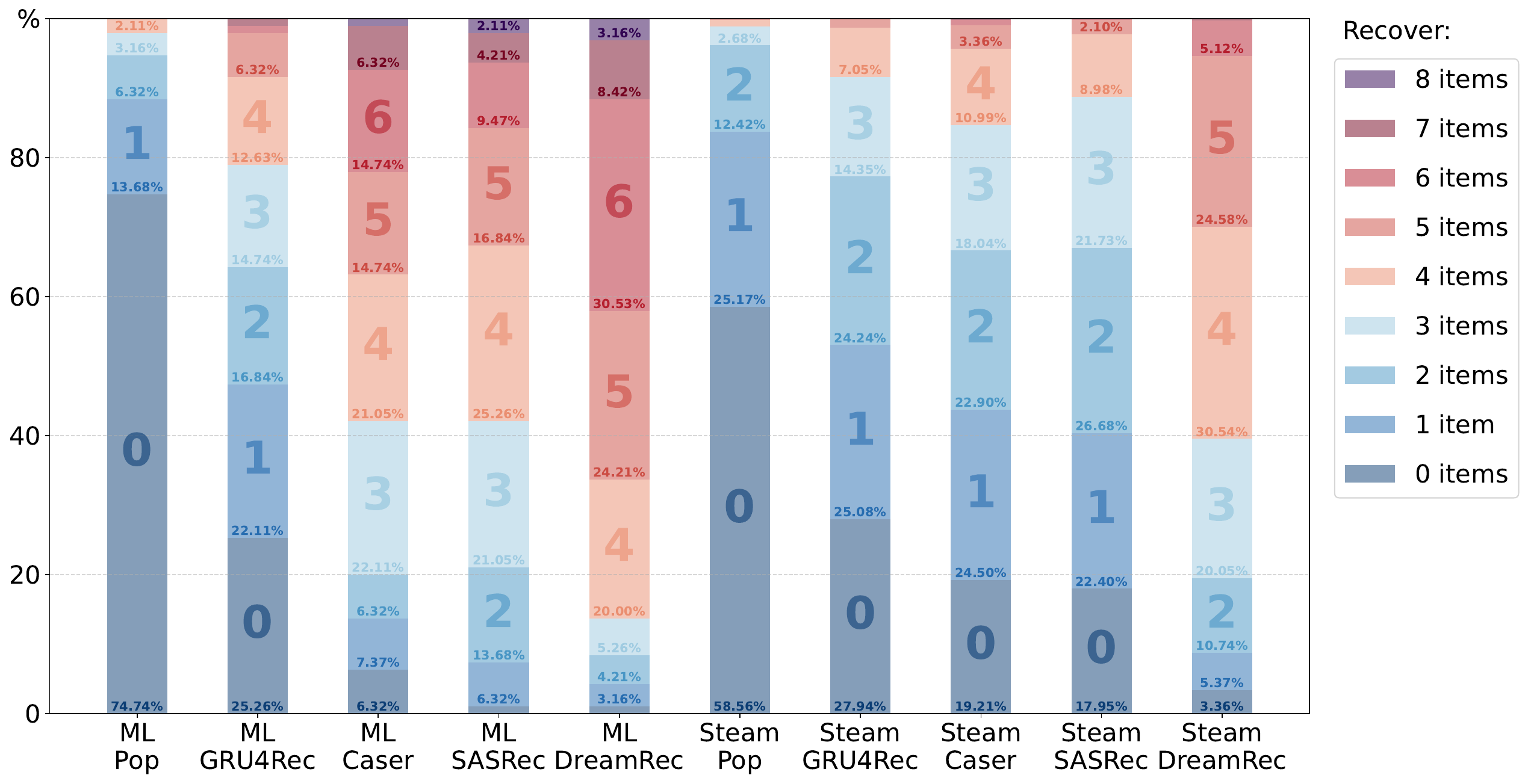}
        \vspace{-0.3cm}
	\caption{The distribution of the number of recovered items on the MovieLens (ML) and Steam datasets, with GRU4Rec, Caser, SASRec, and DreamRec as the sequential recommenders. `Pop' denotes that the 10 most popular movies or games in the training data are provided as the sequence recovery results. The average number of items in the test sequences is 9.99 for the MovieLens dataset, and 8.89 for the Steam dataset.}
	\label{fig:exp_seqnence-recovery}
	\vspace{-0.3cm}
\end{figure*}

\subsection{Experimental Setings}

\subsubsection{Datasets}
We use two datasets from real-world recommendation scenarios: MovieLens and Steam:
\begin{itemize}[leftmargin=*]
    \item\textbf{MovieLens}\footnote{\url{https://grouplens.org/datasets/movielens/}} is a well-known dataset for movie recommendation, containing users' rating history. We preserve titles as the textual descriptions of movies.
    \item \textbf{Steam} \cite{SASRec} dataset contains user reviews of video games on the Steam Store. The titles of video games are also available as textual descriptions.
\end{itemize}
Since tuning the projection layer requires backpropagation from LLaMA, the training phase is more time-consuming than conventional recommenders, and the size of datasets should not be too large.
Therefore, we select the MovieLens100K dataset in our experiment.
For the Steam dataset, we first remove users who have less than 20 reviews, which keeps the same as the processing of MovieLens.
Then we sample 1/3 of users and 1/3 of games and preserve their interactions to acquire a moderate size of dataset.

For both datasets, we first sort all sequences in chronological order and then split the data into training, validation, and testing data at the ratio of 8:1:1.
This splitting strategy ensures that later interactions would not appear in the training set, avoiding any potential of information leakage \cite{data-leakage}.
The statistics of datasets are illustrated in Table \ref{tab: dataset}.

\subsubsection{Implementation details}
We implement all approaches with Python 3.10, PyTorch 2.0.0, and transformers 4.28.0 in a single Nvidia GeForce A40.
We preserve the last 10 interactions as the historical sequence. For sequences with less than 10 interactions, we would pad them to 10 with a padding token.

We first train the sequential recommenders (GRU4Rec \cite{GRU4Rec}, Caser \cite{Caser}, SASRec \cite{SASRec} and DreamRec \cite{dreamrec}) on the training datasets.
We use Adam optimizer, the learning rate is tuned as 0.001 and the batch size is set as 256.
We adopt L2 regularization for all models other than DreamRec and the coefficient is searched in [1e-3, 1e-4, 1e-5, 1e-6, 1e-7], since DreamRec does not require L2 regularization \cite{dreamrec}.
The embedding size is searched in [16, 64, 256, 1024].
The sequential recommenders would be frozen after training.

We would utilize the frozen encoders in the pre-trained sequential recommenders to obtain the hidden representations of interaction sequences.
Let $L$ and $D$ denote the length of the sequences and the dimension of the item embeddings respectively.
For Caser, the size of the hidden representation is $ 1 \times (D + n_f \times s_f)$, where $n_f$ and $s_f$ are the number and size of convolutional kernels respectively \cite{Caser}, we directly employ a linear layer to transfer the hidden representation to be the size of token embedding of LLaMA.
For GRU4Rec, SASRec, and DreamRec, they adopt sequence-to-sequence models (RNN or Transformer encoder) as sequence encoders, and the size of hidden representations is $L \times D$ \cite{GRU4Rec, SASRec}.
Therefore, we first acquire the linear combination of the hidden representations by employing a convolutional filter of size $L \times 1$ to acquire a $1 \times D$ representation, and then adopt the linear projection similar to Caser.

In the training phase of RecInterpreter, we adopt a warmup learning rate schedule: the learning is set as 0.0001 at the 1st epoch, increases linearly to 0.0005 at the 5th epoch, and remains unchanged.
We search the L2 regularization coefficient in the range of [1e-4, 1e-5, 1e-6].
We select LLaMA-7B \cite{llama} as the LLM in our experiment.
For the sequence recovery framework, we set the maximum generated token length as 100, since each hidden representation contains several items.
For the sequence residual framework, we set the maximum generated token length as 50, since the residual item contains only one item.
It takes about 2 hours and 6 hours to train the model for a single epoch in the MovieLens dataset and Steam dataset respectively.
And training for 20 epochs is generally enough for convergence.

\begin{table}[t]
\caption{Statistics of datasets.}
\label{tab: dataset}
\renewcommand\arraystretch{1.1}
\begin{tabular}{ccc}
\hline
Dataset&MovieLens&Steam \\ \hline
\#sequences& 943& 11,938\\
\#items&1,682& 3,581\\
\#interactions&100,000& 274,726\\ \hline

\end{tabular}
\end{table}

\subsection{Sequence-Recovery Result}
\begin{figure}[t]
	\centering
	\small	
	\includegraphics[width=0.47\textwidth, trim={0 1cm 0 1cm}]{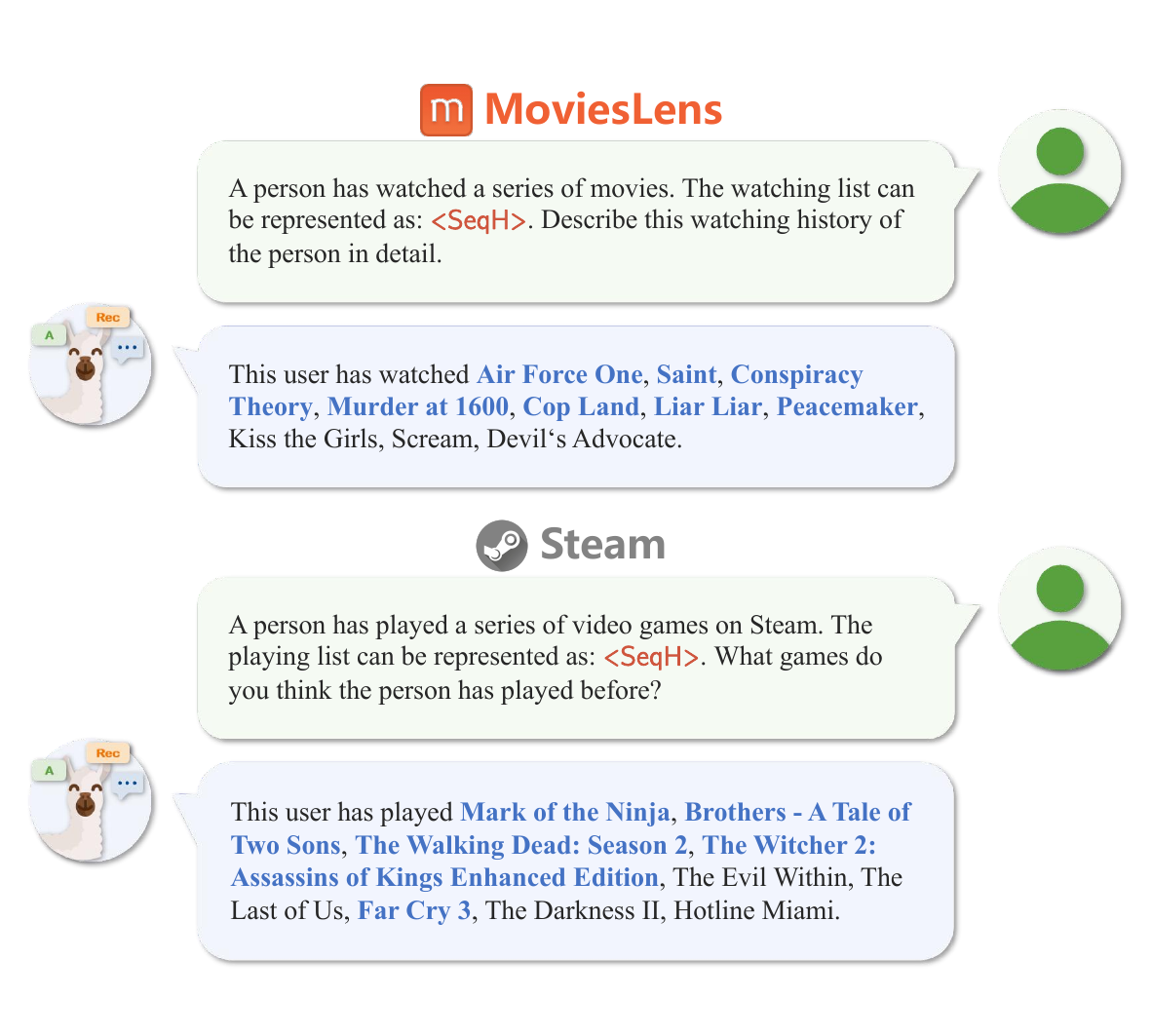}
	\caption{Two cases of the sequence recovery task in MovieLens and Steam datasets. \seqemb{\texttt{<SeqH>}} denotes the hidden representation of the interaction sequence after the projection layer. The \textcolor{myblue2}{blue text} in the response denotes the correctly recovered movies and games from only the hidden representation.}
	\label{fig:exp-case-recovery}
\end{figure}

The straightforward approach to show whether LLaMA could understand the hidden representations of sequential recommenders is to let LLaMA recover the items encoded in the hidden representations with textual descriptions.
In the testing data, each interaction sequence contains 10 movies or video games, and the sequence recovery task is to recover these items based on the hidden representations of the sequential recommenders.
It is worth noting that sequence recovery is not a trivial task, since each dataset contains thousands of items and the hidden representation is highly compressed.
We illustrate the result in Figure \ref{fig:exp_seqnence-recovery}, and we present the observed cases from the MovieLens and Steam datasets in the inference phase in Figure \ref{fig:exp-case-recovery}.

From Figure \ref{fig:exp_seqnence-recovery} we can observe that:
\begin{itemize}[leftmargin=*]
    \item If we naively provide the most popular items in the training data as the recovery of interaction sequences, they can hardly match the real interacted items.
    In the MovieLens dataset, among 74.74\% of the test samples, the popularity-based recovery strategy can not recover any items.
    Similarly, in the Steam dataset, the ratio of recovering 0 items based on popularity is 58.56\%.
    These results suggest that the sequence recovery task is quite challenging, which a simple heuristic of popularity can hardly handle. 
    \item In general, with our designed sequence recovery framework, LLaMA shows the capability of understanding hidden representations of sequential recommenders.
    Notably, in the MovieLens dataset, LLaMA could recover more than 5 items from the hidden representations of Caser, SASRec, and DreamRec for over 35\% of all test samples, and the percentage of recovering more than 3 items from the hidden representations could reach 80\%.
    Therefore, we could safely draw the conclusion that the hidden representations of interaction sequences encoded by sequential recommender are also perceivable for LLaMA, just as the hidden representations of images, audios, and videos.
    \item In the comparison of the MovieLens and Steam datasets, we can observe that LLaMA shows a better understanding of the MovieLens data than the Steam data. One reason is that the game titles in the Steam dataset are more complex than the movie titles in the MovieLens dataset.
    Specifically, the movie titles in the MovieLens dataset are quite simple and clear containing only English words. 
    However, the game titles in the Steam dataset are more complicated. Plenty of the game titles contain the version or provider information, such as \textit{``Swords and Sorcery - Underworld - Definitive Edition''}. Besides, some of the game titles contain other languages than English, such as Chinese or Japanese.
    Therefore, it might be harder for LLaMA to understand the Steam dataset.
    \item In the comparison of different sequential recommenders, we can observe that LLaMA could interpret the hidden representations of DreamREc most precisely. The reason can be that DreamRec adopts the diffusion model for generative recommendation, and diffusion has shown remarkable performance in generation tasks \cite{stable-diffusion, DiffLM}.
    Among the other three sequential recommenders, LLaMA can better understand Caser, SASRec than GRU4Rec. 
    The reason comes from their different model architectures.
    Specifically, Caser adopts CNN to capture the sequential patterns, and CNN has displayed the impressive capability of encoding the global information \cite{resnet}.
    The Transformer encoder employed by SASRec is one of the widely adopted sequence-to-sequence architectures \cite{attn}, which is also the fundamental component in the framework of LLaMA \cite{llama}.
    However, RNN may suffer from issues such as vanishing gradient and exploding gradient \cite{rnn-bad}, thus presenting obstacles for  LLaMA to well understand its hidden representations.
\end{itemize}

\subsection{Sequence-Residual Result}
We have shown that LLaMA could understand the interactions from the hidden representation, but it is hard to recover all items, since the number of candidates is very large and the hidden representation is highly compressed.
Therefore, we design the sequence-residual prompting framework, \ie inspiring LLaAM to identify the residual item by comparing the representations before and after the sequence incorporates the said residual.

\begin{table}[t]
\caption{The result of sequence residual on the MovieLens and Steam datasets. We calculate the accuracy of correctly identifying the residual item in the test data as the evaluation metrics.}
\label{tab: exp-sequence-residual}
\renewcommand\arraystretch{1.1}
\begin{tabular}{ccccc}
\hline
Dataset&GRU4Rec&Caser&SASRec&DreamRec \\ \hline
MovieLens& 52.63\%& 78.95\%&93.68\%&97.89\%\\ \hline
Steam& 17.11\%& 55.03\%&52.60\%&86.33\%\\ \hline
\end{tabular}
\end{table}

We illustrate the result in Table \ref{tab: exp-sequence-residual}, from which we can observe:
\begin{itemize}[leftmargin=*]
    \item LLaMA could identify the residual item based on the hidden representations of DreamRec with high accuracy (97.89\% at the MovieLens dataset and 86.33\% at the Steam dataset.), which verifies the effectiveness of the proposed sequence-residual framework.
    Besides, the remarkable accuracy also suggests that DreamRec could better encode the interaction sequences, owing to the superiority of the diffusion model.
    Among the conventional recommenders, the accuracy of residual item identification is higher based on Caser and SASRec than GRU4Rec.
    The reason is similar to the analysis in the sequence-recovery framework, that Caser and SASRec employ more effective model architectures (CNN and Transformer encoder) than GRU4Rec (RNN).
    \item LLaMA also understands the MovieLens dataset better than the Steam data \wrt the sequence-residual task.
    The reason comes from that the titles of video games are more complicated than the titles of movies.
    However, DreramRec can also be well understood by LLaMA under the sequence-residual framework (with the accuracy of 86.33\%), which further indicates that the complicated game titles are also distinguishable with the hidden representations learned by DreamRec.    
\end{itemize}

\subsection{Instantiate Oracle items}
Since our sequence residual framework could identify the residual item from hidden representations before and after the sequence incorporates the residual item, we could replace the residual item with the generated oracle item by DreamRec.
Accordingly, our RecInterpreter could provide a text description of the oracle item, and instantiate it as the recommendation results.

Similar to the arguments in DreamRec \cite{dreamrec} that the generated oracle items are not limited in the candidate set, we also discover that the instantiations of the oracle items through our RecInterpreter may not exist in the item set of the datasets.
Specifically, we empirically find that 28.13\% of the instantiations are beyond the movies in the MovieLens dataset, and 48.67\% of the instantiations are beyond the video games in the Steam dataset.
Therefore, it is hard to evaluate the performance of DreamRec based on the instantiation of RecInterpreter with traditional evaluation metrics such as Hit Ratio (HR) or normalized discounted cumulative gain (NDCG) \cite{SASRec}.
To this end, we further conduct the evaluation with the assistance of ChatGPT.

Specifically, given an interaction sequence of a user, SASRec could assign preference scores for the candidate items, and select the item with the highest score as the recommendation result.
DreamRec could -generate an oracle item in the form of vector representation, which can be instantiated through the proposed RecInterpreter. 
Afterward, we could ask ChatGPT which item the user prefers among the recommended items provided by SASRec and DreamRec together with a randomly sampled item from the candidate set.
An example of the prompts in the MovieLens dataset shows as follows:
\begin{tcolorbox}[boxrule=0.5pt, , left=5pt, right=5pt, title={ChatGPT Evaluation Prompt Example}]
``A person has watched a series of movies: \texttt{<Watching History>}. Which of the following movies does this person prefer? \texttt{<Movie1>}, \texttt{<Movie2>}, or \texttt{<Movie3>}. Please pick one.''
\end{tcolorbox}
\noindent where \texttt{<Watching History>} is the movie titles in the interactions, and \texttt{<Movie1>}, \texttt{<Movie2>} and \texttt{<Movie3>} are the movie titles of the three recommendation result.

The results are illustrated in Table \ref{tab: exp-instantiate-oracle}. 
We can observe that under the evaluation of ChatGPT, DreamRec outperforms SASRec in the MovieLens dataset, and achieves comparable performance with SASRec in the Steam dataset.
Besides, the learned recommenders SASRec and DreamRec both outperform the naive baseline of random sampling.
This reasonable result suggests that RecInterpreter provides an effective approach to
instantiate the generated oracle items of DreamRec, which completes the full promise of generative recommenders to provide recommendations beyond the constraint of candidate items.

\begin{table}[t]
\caption{The distribution of ChatGPT's selection from the random strategy, SASRec, and DreamRec.}
\label{tab: exp-instantiate-oracle}
\renewcommand\arraystretch{1.1}
\begin{tabular}{cccc}
\hline
Dataset&Random&SASRec&DreamRec \\ \hline
MovieLens& 13.68\%& 35.79\%&50.53\%\\ \hline
Steam& 0.76\%& 51.15\%&48.09\%\\ \hline
\end{tabular}
\end{table}

%% file: chapters/5_conclusion.tex
\section{Conclusion and Limitations}
We propose RecInterpreter, inspiring LLMs to understand the hidden representation of conventional sequential recommenders.
RecInterpreter draws inspiration from recent advances in multi-modal language models, that hidden representation of modality-specific encoders, such as image encoders and audio encoders, could be perceived by LLMs through simple projection.
Therefore, RecInterpreter designs the sequence-recovery and sequence-residual promptings, allowing for LLaMA to understand the hidden representation of sequential recommenders.
Besides, RecInterpreter provides a novel scheme to instantiate the generated oracle items, completing the full promise of generative recommendation.

Meanwhile, RecInterpreter also has a few limitations: 1) the projection is simply set as a linear layer; and 2) the size of the datasets is not large enough.
We believe these can be resolved in further research with more advanced adapters such as Q-former \cite{blip-2}, and larger datasets, with sufficient computation resources.
Moreover, as an initial attempt to inspire LLMs to understand the hidden representation of recommenders, RecInterpreter provides many research opportunities.
For example, online service providers could apply RecInterpreter to their own recommenders, and explore the application of LLMs in their platforms from this perspective. 
Besides, researchers can design other prompting frameworks other than the sequence-recovery and sequence-residual, to better guide LLMs to understand sequential recommenders.
Moreover, exploring the understanding of other recommenders with LLMs, such as collaborative filtering models and conversational recommenders, can also serve as a promising research direction.



